\documentclass[twocolumn]{aastex61}

\newcommand{\bjdtdb}{\ensuremath{\rm {BJD_{TDB}}}}
\newcommand{\feh}{\ensuremath{\left[{\rm Fe}/{\rm H}\right]}}
\newcommand{\teff}{\ensuremath{T_{\rm eff}}}

\newcommand{\msun}{\ensuremath{\,M_\Sun}}
\newcommand{\rsun}{\ensuremath{\,R_\Sun}}
\newcommand{\lsun}{\ensuremath{\,L_\Sun}}
\newcommand{\mj}{\ensuremath{\,M_{\rm J}}}

\newcommand{\rj}{\ensuremath{\,R_{\rm J}}}
\newcommand{\fave}{\langle F \rangle}
\newcommand{\fluxcgs}{10$^9$ erg s$^{-1}$ cm$^{-2}$}

\received{}
\revised{}
\accepted{}
\shorttitle{PHOTOMETRIC FOLLOW-UP OBSERVATIONS OF HAT-P-33b}
\shortauthors{WANG ET AL.}

\begin{document}

\title{Transiting Exoplanet Monitoring Project (TEMP). II. Refined System Parameters and Transit Timing Analysis of HAT-P-33b}

\correspondingauthor{Zhenyu Wu}  
\email{zywu@bao.ac.cn}

\author{Yong-Hao Wang}
\affiliation{Key Laboratory of Optical Astronomy, National Astronomical Observatories, Chinese Academy of Sciences, Beijing 100012, China}
\affiliation{University of Chinese Academy of Sciences, Beijing 10039, China}

\author{Songhu Wang}
\affiliation{Department of Astronomy, Yale University, New Haven, CT 06511, USA}

\author{Hui-Gen Liu}
\affiliation{School of Astronomy and Space Science and Key Laboratory of Modern Astronomy and Astrophysics in Ministry of Education, Nanjing University, Nanjing 210093, China}

\author{Tobias C. Hinse}
\affiliation{Korea Astronomy and Space Science Institute, Daejeon 305-348, Republic of Korea}
\affiliation{Armagh Observatory, Armagh BT61 9DG, Northern Ireland, United Kingdom}

\author{Gregory Laughlin}
\affiliation{Department of Astronomy, Yale University, New Haven, CT 06511, USA}

\author{Dong-Hong Wu}
\affiliation{School of Astronomy and Space Science and Key Laboratory of Modern Astronomy and Astrophysics in Ministry of Education, Nanjing University, Nanjing 210093, China}

\author{Xiaojia Zhang}
\affiliation{Key Laboratory of Optical Astronomy, National Astronomical Observatories, Chinese Academy of Sciences, Beijing 100012, China}
\affiliation{Department of Physics, Center for Astrophysics and Institute for Advanced Studies, Tsinghua University, Beijing 100086, China}

\author{Xu Zhou}
\affiliation{Key Laboratory of Optical Astronomy, National Astronomical Observatories, Chinese Academy of Sciences, Beijing 100012, China}

\author{Zhenyu Wu}
\affiliation{Key Laboratory of Optical Astronomy, National Astronomical Observatories, Chinese Academy of Sciences, Beijing 100012, China}
\affiliation{University of Chinese Academy of Sciences, Beijing 10039, China}

\author{Ji-Lin Zhou}
\affiliation{School of Astronomy and Space Science and Key Laboratory of Modern Astronomy and Astrophysics in Ministry of Education, Nanjing University, Nanjing 210093, China}

\author{R. A. Wittenmyer}
\affiliation{Computational Engineering and Science Research Centre, University of Southern Queensland, Toowoomba, Queensland 4350, Australia}
\affiliation{School of Physics and Australian Centre for Astrobiology, UNSW Australia, Sydney 2052, Australia}

\author{Jason Eastman}
\affiliation{Harvard-Smithsonian Center for Astrophysics, Cambridge, MA 02138, USA}

\author{Hui Zhang}
\affiliation{School of Astronomy and Space Science and Key Laboratory of Modern Astronomy and Astrophysics in Ministry of Education, Nanjing University, Nanjing 210093, China}

\author{Yasunori Hori}
\affiliation{Astrobiology Center, NINS, 2-21-1 Osawa, Mitaka, Tokyo 181-8588, Japan}
\affiliation{National Astronomical Observatory of Japan, NINS, 2-21-1 Osawa, Mitaka, Tokyo 181-8588, Japan}

\author{Norio Narita}
\affiliation{Astrobiology Center, NINS, 2-21-1 Osawa, Mitaka, Tokyo 181-8588, Japan}
\affiliation{National Astronomical Observatory of Japan, NINS, 2-21-1 Osawa, Mitaka, Tokyo 181-8588, Japan}
\affiliation{Department of Astronomy, The University of Tokyo, 7-3-1 Hongo, Bunkyo-ku, Tokyo 113-0033, Japan}

\author{Yuanyuan Chen}
\affiliation{Purple Mountain Observatory, Chinese Academy of Sciences, Nanjing 210008, China}

\author{Jun Ma}
\affiliation{Key Laboratory of Optical Astronomy, National Astronomical Observatories, Chinese Academy of Sciences, Beijing 100012, China}
\affiliation{University of Chinese Academy of Sciences, Beijing 10039, China}

\author{Xiyan Peng}
\affiliation{Key Laboratory of Optical Astronomy, National Astronomical Observatories, Chinese Academy of Sciences, Beijing 100012, China}

\author{Tian-Meng Zhang}
\affiliation{Key Laboratory of Optical Astronomy, National Astronomical Observatories, Chinese Academy of Sciences, Beijing 100012, China}

\author{Hu Zou}
\affiliation{Key Laboratory of Optical Astronomy, National Astronomical Observatories, Chinese Academy of Sciences, Beijing 100012, China}

\author{Jun-Dan Nie}
\affiliation{Key Laboratory of Optical Astronomy, National Astronomical Observatories, Chinese Academy of Sciences, Beijing 100012, China}

\author{Zhi-Min Zhou}
\affiliation{Key Laboratory of Optical Astronomy, National Astronomical Observatories, Chinese Academy of Sciences, Beijing 100012, China}

\begin{abstract}

We present ten $R$-band photometric observations of eight different transits of the hot Jupiter HAT-P-33b, which has been targeted by our Transiting Exoplanet Monitoring Project (TEMP). The data were obtained by two telescopes at the Xinglong Station of National Astronomical Observatories of China (NAOC) from 2013 December through 2016 January, and exhibit photometric scatter of $1.6-3.0\,\rm{mmag}$. After jointly analyzing the previously published photometric data, radial-velocity (RV) measurements, and our new light curves, we revisit the system parameters and orbital ephemeris for the HAT-P-33b system. Our results are consistent with the published values except for the planet-to-star radius ratio ($R_{P}/R_{*}$), the ingress/egress duration ($\tau$) and the total duration ($T_{14}$), which together indicate a slightly shallower and shorter transit shape. Our results are based on more complete light curves, whereas the previously published work had only one complete transit light curve. No significant anomalies in Transit Timing Variations (TTVs) are found, and we place upper mass limits on potential perturbers, largely supplanting the loose constraints provided by the extant RV data. The TTV limits are stronger near mean-motion resonances, especially for the low-order commensurabilities. We can exclude the existence of a perturber with mass larger than 0.6, 0.3, 0.5, 0.5, and $0.3\,{\rm M_\oplus}$ near the 1:3, 1:2, 2:3, 3:2, and 2:1 resonances, respectively.

\end{abstract}


\section{INTRODUCTION} \label{sec1}

As the length of the catalog mounts\setcounter{footnote}{14}\footnote{$\,$See http://exoplanets.org/ for a list of confirmed exoplanets.}, so to does the importance of characterizing the alien worlds. A better understand the extrasolar planets' compositions, their formation and their evolution constitutes a grand challenge for the twenty-first century. With these larger goals as a motivation, we initialized the Transit Exoplanets Monitoring Project (TEMP) to specifically study the transiting exoplanet systems with high-precision photometric follow-up observations \citep{wang2016}. 

High-precision photometric follow-ups lead to more accurate measurements of planetary radii and orbital inclinations, and combining with the RV method permits determinations of planetary masses which in turn give densities, and hence the planetary compositions \citep{s2005}. With improved photometry, we can determine more precise orbital ephemerides, which streamline future research studies, including those that draw on the Rossiter-McLaughlin (RM) effect \citep{n2011, s2011, s2013}, transmission spectra \citep{man2016} and spectroscopy at secondary eclipse \citep{star2016}. Furthermore, we can perform transit timing variation (TTV) analyse with high-precision photometric data. These provide us the powerful tools to detect close-in companions in known hot Jupiter systems and hence enable the zeroth-order test of competing formation scenarios for hot Jupiters \citep{l1996, b2000, f2008, n2008, w2003, w2011, bll2016}. Moreover, with TTVs in hand, we can confirm the planetary nature and measure masses for planets in multi-transiting systems \citep{lithwick2012, xie2014, hadden2015}. Most of the multi-transiting systems that are currently known were detected by the Kepler space telescope \citep{fab2014}. The host stars, however, of the Kepler-detected systems are mostly too faint for feasible RV follow-up observations from the ground using small to medium aperture optical telescopes, incentivizing the search for TTVs among these systems.

In addition, we can confirm candidate exoplanets and refine their orbital ephemerides through photometric monitoring of planets that have been observed for a limited number of transits in the K2 data sets \citep{h2014} and/or from the forthcoming TESS mission \citep{r2015}. High-precision photometric follow-ups in multi-band also provide a means to determine the chemical compositions and atmospheric properties for exoplanets \citep{m2013, f2013, l2013, s2016}.   

In the first stage of TEMP, we have focused primarily on monitoring the hot Jupiters found by ground-based transiting surveys, concentrating on those for which only limited  photometric follow-up observations have been published. In many cases, the parameters for such systems are both imprecise and incomplete. These targets, therefore, offer an optimal scientific benefit and studies similar to these provided by TEMP have been presented by other groups \citep{becker2015, see2015, co2017}, demonstrating that the TTVs provide a key avenue for insights into exoplanetary formation and evolution. In this paper, we present the scientific results that emerged from a monitoring campaign on HAT-P-33b. This system was chosen because of its large RV residuals and the existence of only sparse data in form of incomplete light curves \citep{Hartman2011}. 

HAT-P-33b was discovered by \citet{Hartman2011}, who found the planet to be a highly inflated hot Jupiter ($M_{P}$=$0.763\, {\rm {\mj}}$, $R_{P}$=$1.827\, {\rm {\rj}}$) transiting a late-F dwarf star ($M_{*}$=$1.403\, {\rm M_{\odot}}$, $R_{*}$=$1.777\, {\rm R_{\odot}}$) with an orbital period of $3.474474\,{\rm days}$. Seven light curves were presented in their work, but only one is complete. Following the discovery by \citet{Hartman2011}, four more RV measurements (obtained also using Keck HIRES) were presented in \citet{knutson2014} bringing the total number of RV observations to be 26. The extended data showed no evidence of long-period companions within the HAT-P-33b system.

In this work, we present ten new light curves of eight different transits of HAT-P-33b. The light curves are all complete, with a typical photometric precision better than $2.0\,{\rm mmag}$ save one, which is partial and has a precision of $3.0\,{\rm mmag}$. Based on our photometric data and the extended RV measurements \citep{knutson2014}, we revisit the system parameters, refine the orbital ephemeris, and explore the possibility of existence of additional planets in the system. 

This paper is organized as follows. We describe the photometric observations and data reduction in \S \ref{sec2}. The data analysis is presented in \S \ref{sec3}. In \S \ref{sec4}, we give the results and discussion. Finally, a brief summary of our work is presented in \S \ref{sec5}.

\section{OBSERVATIONS AND DATA REDUCTION} \label{sec2}

\setlength{\tabcolsep}{1.1pt}
\begin{deluxetable*}{cccccccccccc}[t!]
\tabletypesize{\scriptsize}
\tablewidth{0pt}
\tablecaption{Overview of Observations and Data Reduction}
\tablehead{
\colhead{Date}  &  \colhead{Time}   &   \colhead{Telescope}\tablenotemark{a} & \colhead{Filter} &  \colhead{Frames}  &  \colhead{Exposure}  & \colhead{Read} & \colhead{Airmass} &\colhead{Moon}  &\colhead{Comp.}  & \colhead{Aperture}\tablenotemark{b}  & \colhead{Scatter}\tablenotemark{c} \\
\colhead{(UTC)} &\colhead{(UTC)} & \colhead{} & \colhead{}  &  \colhead{} & \colhead{(second)} & \colhead{(second)} & \colhead{} & \colhead{illum.} &\colhead{Stars}& \colhead{(pixels)}  &  \colhead{(mmag)} }
\startdata
2013 Dec 09 & { }{ }14:40:56-20:05:27 & { } Schmidt &             $R$ &     247  &   60,80    & 4&   $1.41\rightarrow1.01\rightarrow1.07$   & 0.51  &5   &10   & 2.0 \\
2014 Feb 27 & { }{ }10:54:42-18:21:57 & { } $60\,{\rm cm}$ &   $R$ &     373  &   60,70    &3&   $1.13\rightarrow1.01\rightarrow2.01$   & 0.04   &3  &11    &2.0 \\
2014 Feb 27 & { }{ }11:01:37-17:54:05 &  { } Schmidt &             $R$ &    295  &   60,110   &4&   $1.12\rightarrow1.01\rightarrow1.75$   &  0.04  &3  &14    &1.9\\ 
2014 Mar 06 & { }{ }10:48:53-18:02:13 & { }$60\,{\rm cm}$&     $R$ &     397  &   60         &3&   $1.09\rightarrow1.01\rightarrow2.10$   &0.32    &2   &10   &1.6  \\
2014 Mar 06 & { }{ }11:09:38-17:27:35 & { } Schmidt  &            $R$ &     251  &   80,110  &4&    $1.06\rightarrow1.01\rightarrow1.76$   & 0.32   &2   &14   &1.7 \\
2015 Jan 16 & { }{ }14:52:35-20:47:16 &  { } Schmidt   &          $R$ &     343   &   55        &4&   $1.03\rightarrow1.01\rightarrow1.81$   &  0.19   &4   &14   &1.7 \\
2015 Jan 23 & { }{ }13:23:02-19:45:43 & { } Schmidt    &          $R$ &     322   &   52        &4&    $1.12\rightarrow1.01\rightarrow1.56$   &  0.14  &4    & 11 &1.9\\
2015 Jan 30 & { }{ }12:23:56-18:56:37 & { } Schmidt  &            $R$ &     465   &   40,50   &4&   $1.18\rightarrow1.01\rightarrow1.44$   & 0.84     &4   &10  &1.9  \\
2015 Feb 13 & { }{ }10:40:54-16:35:31 &  { } Schmidt  &          $R$ &     353   &   45,55   &4&   $1.34\rightarrow1.01\rightarrow1.15$    &0.34     &2   &13   &1.8  \\
2016 Jan 09 & { }{ }12:16:06-15:15:30 &  { }$60\,{\rm cm}$&    $R$ &    695   &   10         &3&       $1.55\rightarrow1.04$    &0.00     &3   &10   &3.0  \\
\enddata
\tablenotetext{a}{The telescopes belong to the Xinglong Station operated by National Astronomical Observatories of China (NAOC).}
\tablenotetext{b}{The aperture indicates the aperture diameter around stars.}
\tablenotetext{c}{The scatter indicates the RMS of residuals from our best-fitting model.}
\label{table1}
\end{deluxetable*}

We have recorded a total of ten light curves of eight different transits events observed by two telescopes (a $60/90\,{\rm cm}$ Schmidt and a $60\,{\rm cm}$ telescope) at Xinglong Station operated by National Astronomical Observatories of China (NAOC) between 2013 December and 2016 January. Two of the transit events were observed by the two telescopes simultaneously.

The first seven transits events were monitored by the $60/90\,{\rm cm}$ Schmidt telescope. It has a $4{\rm K} \times 4{\rm K}$ CCD with a $\sim94' \times \sim94'$ field of view, which gives a pixel scale of $1.38''\,{\rm pixel ^{-1}}$ and a typical readout time of $93\,{\rm s}$ \citep{zhou1999, zhou2001}. To reduce the readout times and increase the duty cycle of the observations, the images were windowed down to $512 \times 512$ pixels with $1 \times 1$ binning, resulting in a reduced readout time of $4\,{\rm s}$. 

The transit events that occurred on UT 2014 February 27 and UT 2014 March 6 were also simultaneously observed by the $60\,{\rm cm}$, and the last transit in our sequence was also monitored with this telescope. The $60\,{\rm cm}$ telescope is equipped with a $512 \times 512$ CCD and covers a field of view of $17' \times 17'$, resulting in a pixel scale of $1.95''\,{\rm pixel ^{-1}}$. No windowing with $1 \times 1$ binning was performed during these observations, giving a standard readout time of $3\,{\rm s}$. 

It is common practice to defocus the telescope in order to optimize signal-to-noise and to keep photoelectron counts within the CCD's range of linear response \citep{Southworth2009}. Broadened stellar Point Spread Functions (PSFs) are less sensitive to focus or telescope pointing changes, which would otherwise cause systematic errors. Defocusing produces longer exposure times, which increase the duty cycle of observations and reduce Poisson or scintillation noise \citep{hin2015}. 

In our observations of HAT-P-33b, which is a $V_{mag}=11.19$ star, we slightly defocused our telescopes.
The linear range of the CCD is maintained for target counts less than 30,000. For the sake of conservatism, we keep our target at a typical counts of 20,000 which is reached within 15 seconds in a clear night. Fifteen seconds is too short, however to achieve optimal reduction of Poisson and scintillation noise, which further motivates our defocusing of the CCD images. The background counts is about 300 within our typical 60-second exposure time.  
We adjust exposure times throughout each data-taking session in order to maintain counts that fall within the linear regime of the CCD. The exposure time, however, was kept fixed during the ingress and egress phases to avoid affecting the precision of transit timing, which is a critical aspect of our work. The telescope time stamp server was synchronized on a nightly basis with the US Naval Observatory (USNO) time\setcounter{footnote}{15}\footnote{$\,$http://tycho.usno.navy.mil/.}. Timings are measured accurately to within one second and recorded using the UTC time standard. A summary of our observations is listed in Table~\ref{table1}.

All the data have been calibrated using standard procedure, including overscan correction and flat-fielding for data from the Schmidt telescope, as well as bias correction and flat-fielding for data from the $60\,{\rm cm}$ telescope. Twilight sky flats were obtained by the $60\,{\rm cm}$ telescope, whereas dome flats were taken with the Schmidt. We used SExtractor \citep{Bertin1996} to perform aperture differential photometry. 
All the stars in the field with enough flux were tested for photometric non-variability, and the most favorable sources were chosen as reference stars.
With the reference stars, we obtained the differential light curve which has the smallest root-mean-square (RMS) scatter for each transit, by manually varying the aperture diameter from 8 to 16 pixels. A summary of the aperture photometry is given in Table~\ref{table1}.  We then removed trends that may be caused by the variation of airmass and intrinsic stellar variability, by performing a linear fit to the out-of-transit data. To maintain timing consistency, we converted the UTC time stamps to to Barycentric Julian Date in the TDB time standard (${\rm BJD}_{\rm TDB}$) for each light curve using the online procedure\setcounter{footnote}{16}\footnote{$\,$http://astroutils.astronomy.ohio-state.edu/time/utc2bjd.html.}. The final set of 10 recorded light curves are listed in Table~\ref{table2}, in total, these data comprise 3732 measurements. 

\startlongtable
\setlength{\tabcolsep}{1.1pt}
\begin{deluxetable}{ccccc}
\tablewidth{0pt}
\tablecaption{Photometry of HAT-P-33}
\tablehead{
\colhead{${\rm BJD_{TDB}}$\tablenotemark{a}} & \colhead{Relative Flux} & \colhead{Scatter}  & \colhead{Telescope} &\colhead{Filter}
}
\startdata
  2456636.117240  &   0.9990  &  0.0020 &  Schmidt &  R  \\
  2456636.117981  &   0.9990  &  0.0020 &  Schmidt &  R  \\
  2456636.118722  &   1.0000  &  0.0020 &  Schmidt &  R  \\
  2456636.121361  &   1.0000  &  0.0020 &  Schmidt &  R  \\
  2456636.122877  &   0.9963  &  0.0020 &  Schmidt &  R  \\
  2456636.123618  &   1.0018  &  0.0020 &  Schmidt &  R  \\
  2456636.124359  &   1.0009  &  0.0020 &  Schmidt &  R  \\
  2456636.125099  &   1.0018  &  0.0020 &  Schmidt &  R  \\
  2456636.125840  &   1.0036  &  0.0020 &  Schmidt &  R  \\
  2456636.126569  &   1.0000  &  0.0020 &  Schmidt &  R  \\
  2456636.127310  &   0.9972  &  0.0020 &  Schmidt &  R  \\
  2456636.128051  &   0.9990  &  0.0020 &  Schmidt &  R  \\
  2456636.128803  &   1.0000  &  0.0020 &  Schmidt &  R  \\
  2456636.129521  &   0.9972  &  0.0020 &  Schmidt &  R  \\
  2456636.130273  &   0.9990  &  0.0020 &  Schmidt &  R  \\
\enddata
\tablenotetext{a}{ All the timing throughout the paper are based on ${\rm BJD_{TDB}}$, calculated from Coordinated Universal Time (UTC) using the procedure developed by \citet{eastman2010}.}
\tablecomments{Table 2 is available in its entirety in the machine readable format. A portion is shown here for guidance regarding its form and content.}
\label{table2}
\end{deluxetable}

\section{DATA ANALYSIS} \label{sec3}

We applied the EXOFAST\setcounter{footnote}{17}\footnote{Online procedure to see http://astroutils.astronomy.ohio-state.edu/exofast/exofast.shtml.} (a fast exoplanetary fitting package in IDL) developed by \citet{eastman2013} to perform our data modeling.
The package can simultaneously fit transit and RV data with given priors, robustly deriving the parameter values and their uncertainties using the differential evolution Markov chain Monte Carlo (DE-MC) algorithm. At each Markov chain step, EXOFAST employs the Torres relations to calculate $M_{*}$ and $R_{*}$ with given $\teff$, $\feh$, and $\log(g_*)$ \citep{2010to}.

To revisit the system parameters of HAT-P-33b, we performed a global fit based on our seven light curves and the extended RV measurements from \citet{knutson2014}. The priors of the system parameters used in the fit were obtained from \citet{Hartman2011}, and are presented in Table~\ref{table3}. We also obtained the priors for the limb darkening parameters in the $R$ band ($u_1$=0.2631, $u_2$=0.3155) following the description in \citet{c2011}. 
As a first step, EXOFAST fitted the RV and transit data sets independently and scaled the uncertainties to obtain a reduced $\chi^2_{\rm red}=1$ for each best-fitting model. Then it performed a global fit based on both data sets. A total of 32 simultaneous chains were constructed in our fit, each having a maximum of 100,000 steps. As described in \citet{eastman2013}, the Markov chains are considered to have converged when both the Gelman-Rubin statistic is less than 1.01 and the number of independent draws is greater than 1000 for all parameters. Only after passing this test 6 consecutive times, can the chains be considered well-mixed and EXOFAST will stop. As a final step, we evaluated the well-mixed results to obtain best-fitting values with $1\,\sigma$ error bars for system parameters, which are also listed in Table~\ref{table3}.

To accurately measure the mid-transit times for all seventeen light curves, we separately performed a fit for each light curve in conjunction with the extended RV measurements from \citet{knutson2014}. The time stamps of the published light curves were converted to ${\rm BJD}_{\rm TDB}$ for reasons of consistency. In these fits, we fixed the system parameters to the values obtained from the aforementioned global fit, excepting $T_{\rm c}$ and baseline flux of the light curve ($F_0$), which were allowed to float instead. For the published light curves from \citet{Hartman2011}, the limb darkening parameters were fixed to different values in diverse bands\setcounter{footnote}{18}\footnote{$\,$For the $i$ band, $u_1$=0.1799, $u_2$=0.3748; for the $z$ band, $u_1$=0.1294, $u_2$=0.3656; for the $g$ band, $u_1$=0.4216, $u_2$=0.3278.} during fitting processes. After a fitting process similar to the global fit mentioned above, we had an estimate for the mid-transit time for each transit event. 

\startlongtable
\setlength{\tabcolsep}{1.1pt}
\begin{deluxetable*}{ccccc}
\tablewidth{0pt}
\tablecaption{System Parameters for HAT-P-33}
\tablehead{\colhead{~~~Parameter} & \colhead{Units} & \colhead{This Work} & \colhead{Hartman et al. (2011)} & \colhead{Knutson et al. (2014)}  }
\startdata
\sidehead{Stellar Parameters:}
                           ~~~$M_{*}$\dotfill &Mass (\msun)\dotfill & $1.42_{-0.15}^{+0.16}$          & $1.403 \pm 0.096$                & $1.403 \pm 0.096$\tablenotemark{a} \\
                         ~~~$R_{*}$\dotfill &Radius (\rsun)\dotfill & $1.91_{-0.20}^{+0.26}$            & $1.777 \pm 0.280$                & ... \\
                     ~~~$L_{*}$\dotfill &Luminosity (\lsun)\dotfill & $5.7_{-1.6}^{+2.3}$                    & $4.73_{-1.25}^{+1.87}$       & ...\\
                         ~~~$\rho_*$\dotfill &Density (cgs)\dotfill & $0.289_{-0.081}^{+0.098}$           & ...                                        & ...\\
              ~~~$\log(g_*)$\dotfill &Surface gravity (cgs)\dotfill & $4.030_{-0.090}^{+0.079}$      & $4.09\pm 0.11$                   & $4.09\pm 0.11$\tablenotemark{a} \\             
               ~~~$\teff$\dotfill &Effective temperature (K)\dotfill & $6460_{-290}^{+300}$               & $6401\pm88$                      & $6401\pm88$\tablenotemark{a} \\
               ~~~$\feh$\dotfill &Metalicity\dotfill & $0.01\pm0.31$                                                    & $0.05\pm0.08$                   & $0.05\pm0.08$\tablenotemark{a} \\
\sidehead{Planetary Parameters:}
                               ~~~$e$\dotfill &Eccentricity\dotfill & $0.180_{-0.096}^{+0.11}$                      & $0.148\pm0.081$                & $0.13_{-0.1}^{+0.19}$ \\
    ~~~$\omega_*$\dotfill &Argument of periastron (degrees)\dotfill & $88_{-34}^{+33}$              & $96\pm119$                        & $15\pm22$ \\
                              ~~~$P$\dotfill &Period (days)\dotfill & $3.47447472\pm0.00000088$\tablenotemark{b}     & $3.474474\pm0.000001$      & ...  \\
                       ~~~$a$\dotfill &Semi-major axis (AU)\dotfill & $0.0505\pm0.0018$                        & $0.0503\pm0.0011$           & ...\\
                             ~~~$M_{P}$\dotfill &Mass (\mj)\dotfill & $0.72_{-0.12}^{+0.13}$                       & $0.763\pm0.117$               & $0.65\pm0.14$\\
                           ~~~$R_{P}$\dotfill &Radius (\rj)\dotfill & $1.87_{-0.20}^{+0.26}$                         & $1.827\pm0.290$               & ...\\
                       ~~~$\rho_{P}$\dotfill &Density (cgs)\dotfill & $0.134_{-0.042}^{+0.053}$                & $0.15_{-0.05}^{+0.11}$       &...\\
                  ~~~$\log(g_{P})$\dotfill &Surface gravity\dotfill & $2.70_{-0.11}^{+0.10}$                     & $2.75\pm0.13$                   &...\\
           ~~~$T_{eq}$\dotfill &Equilibrium Temperature (K)\dotfill & $1920_{-120}^{+140}$               & $1838\pm133$                  &...\\
                       ~~~$\Theta$\dotfill &Safronov Number\dotfill & $0.0271_{-0.0050}^{+0.0056}$       & $0.030_{-0.007}^{+0.005}$    &...\\
               ~~~$\fave$\dotfill &Incident flux (\fluxcgs)\dotfill & $2.96_{-0.65}^{+0.84}$                       & $2.58_{-0.61}^{+0.93}$         &...\\
\sidehead{RV Parameters:}
                              ~~~$e\cos\omega_*$\dotfill & \dotfill & $0.004_{-0.086}^{+0.092}$                           & $0.040\pm0.078$            & $0.114_{-0.097}^{+0.16}$ \\
                              ~~~$e\sin\omega_*$\dotfill & \dotfill & $0.154_{-0.096}^{+0.11}$                               & $0.073\pm0.138$            & $0.015_{-0.023}^{+0.13}$ \\
           ~~~$T_{P}$\dotfill &Time of periastron (\bjdtdb)\dotfill & $2457046.20_{-0.23}^{+0.22}$              & ...                                     & ... \\
                    ~~~$K$\dotfill &RV semi-amplitude (m/s)\dotfill & $78\pm12$                                              & $82.8\pm12.0$                & $72_{-16}^{+19}$\\
                 ~~~$M_P\sin i$\dotfill &Minimum mass (\mj)\dotfill & $0.72_{-0.12}^{+0.13}$                         & ...                                     & ...\\
                       ~~~$M_{P}/M_{*}$\dotfill &Mass ratio\dotfill & $0.000484_{-0.000076}^{+0.000077}$       & ...                                     & ...\\ 
               ~~~$\gamma$\dotfill &Systemic velocity (m/s)\dotfill & $-7\pm11$                                              & ...                                    & ...\\
              ~~~$\dot{\gamma}$\dotfill &RV slope (m/s/day)\dotfill & $-0.024\pm0.018$                                & ...                                     & $-0.021_{-0.023}^{+0.02}$\\
\sidehead{Primary Transit Parameters:}
                ~~~$T_C$\dotfill &Time of transit (\bjdtdb)\dotfill & $2456035.137750\pm0.000272$\tablenotemark{b}  { }   & $2455100.50255\pm0.00023$   & ...\\
~~~$R_{P}/R_{*}$\dotfill &Radius of planet in stellar radii\dotfill & $0.10097_{-0.00052}^{+0.00056}$          & $0.1057\pm0.0011$                & ...\\
     ~~~$a/R_{*}$\dotfill &Semi-major axis in stellar radii\dotfill & $5.69_{-0.59}^{+0.58}$                              & $6.08_{-0.72}^{+0.98}$            & ...\\
              ~~~$u_1$\dotfill &linear limb-darkening coeff\dotfill & $0.264\pm0.026$                                         & ...                                              & ...\\
              ~~~$u_2$\dotfill &quadratic limb-darkening coeff\dotfill & $0.315\pm0.037$                                      & ...                                              & ...\\
                      ~~~$i$\dotfill &Inclination (degrees)\dotfill & $88.2_{-1.3}^{+1.2}$                                         & $86.7_{-1.2}^{+0.8}$               &  $86.7_{-1.2}^{+0.8}$\tablenotemark{a} \\                                                 
                      ~~~$b$\dotfill &Impact Parameter\dotfill & $0.151_{-0.098}^{+0.10}$                                   & $0.325\pm0.002$                          & ...\\
                          ~~~$\delta$\dotfill &Transit depth\dotfill & $0.01020\pm0.00011$                                            & ...                                           & ...\\
                ~~~$T_{FWHM}$\dotfill &FWHM duration (days)\dotfill & $0.16354_{-0.00072}^{+0.00070}$          & ...                                            & ...\\
          ~~~$\tau$\dotfill &Ingress/egress duration (days)\dotfill & $0.01707_{-0.00036}^{+0.00080}$                &  $0.0194\pm0.0002$              & ...\\ 
           ~~~$T_{14}$\dotfill &Total duration (days)\dotfill & $0.18075_{-0.00089}^{+0.00097}$                           &   $0.1836\pm0.0007$         & ...\\
      ~~~$P_{T}$\dotfill &A priori non-grazing transit prob\dotfill & $0.188_{-0.035}^{+0.054}$                            & ...                                            & ...\\ 
                ~~~$P_{T,G}$\dotfill &A priori transit prob\dotfill & $0.231_{-0.043}^{+0.066}$                                   & ...                                           & ...\\
                            ~~~$F_0$\dotfill &Baseline flux\dotfill & $0.999837\pm0.000061$                                          & ...                                           & ...\\
\sidehead{Secondary Eclipse Parameters:}
              ~~~$T_{S}$\dotfill &Time of eclipse (\bjdtdb)\dotfill & $2457044.48_{-0.19}^{+0.21}$                           & $2455102.330\pm0.175$      & ...\\
                       ~~~$b_{S}$\dotfill &Impact parameter\dotfill & $0.21_{-0.13}^{+0.14}$                                        & ...                                           & ...\\
              ~~~$T_{S,FWHM}$\dotfill &FWHM duration (days)\dotfill & $0.219_{-0.037}^{+0.053}$                      & ...                                            & ...\\
        ~~~$\tau_S$\dotfill &Ingress/egress duration (days)\dotfill & $0.0239_{-0.0044}^{+0.0064}$                     & $0.0230\pm0.0085$              & ...\\ 
               ~~~$T_{S,14}$\dotfill &Total duration (da2ys)\dotfill & $0.243_{-0.041}^{+0.059}$                                 & $0.2090\pm0.0480$              & ...\\
      ~~~$P_{S}$\dotfill &A priori non-grazing eclipse prob\dotfill & $0.1379_{-0.0031}^{+0.0046}$                      & ...                                           & ...\\
                ~~~$P_{S,G}$\dotfill &A priori eclipse prob\dotfill & $0.1689_{-0.0038}^{+0.0057}$                            & ...                                           & ... \\
\enddata
\tablenotetext{a}{In \citet{knutson2014}, the stellar parameters (including $M_{*}$, $\log(g_*)$, $\teff$, $\feh$) and orbital inclination ($i$) were adopted from \citet{Hartman2011}.}
\tablenotetext{b}{We got $P$ and $T_C$ through a linear fit based on the mid-transit times which are calculated from the new and published light curves (see S\S \ref{sec4.2}).}
\label{table3}
\begin{flushleft}
\tablecomments{The published system parameters of HAT-P-33 from the literatures \citep{Hartman2011, knutson2014} are presented for comparison.} 
\end{flushleft}
\end{deluxetable*}
\clearpage

\section{RESULT AND DISCUSSION} \label{sec4}

\subsection{System Parameters} \label{sec4.1}

\begin{figure}[b]
\epsscale{0.85}
\plotone{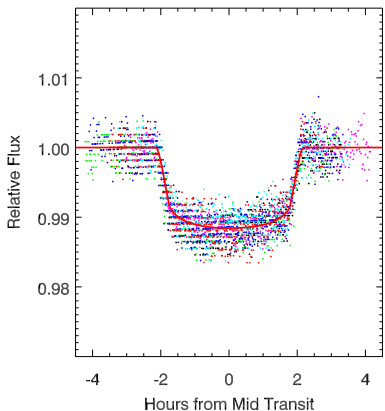}
\caption{Phased light curve of HAT-P-33b transits with different colors representing different light curves. To revisit the system parameters, seven light curves were simultaneously fitted with the published RV observations (Figure~\ref{fig2}) as described in \S \ref{sec3}, resulting in the best-fitting model shown by the solid red line. 
\label{fig1}}
\end{figure}

\begin{figure}[b]
\epsscale{0.85}
\plotone{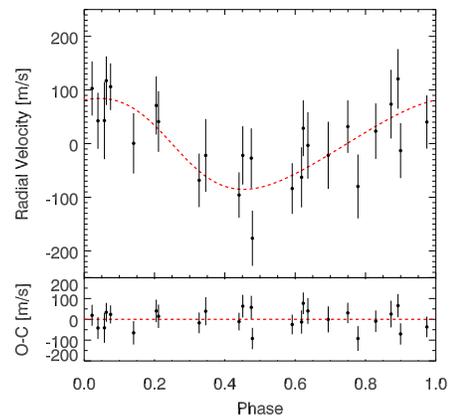}
\caption{The radial velocity observations of HAT-P-33 from \citet{knutson2014}, jointly fitted with our photometric data (see Figure~\ref{fig1} and \S \ref{sec3}), resulting in the best-fitting keplerian orbit model shown by the dashed red line. The residuals from the best-fitting model with an RMS scatter of $47.5\,\rm{m\,s^{-1}}$ is shown at the bottom.
\label{fig2}}
\end{figure}

As the result of global fit, the final parameters for HAT-P-33 system together with the results from previous work \citep{Hartman2011, knutson2014} are listed in Table~\ref{table3}. The resulting best-fitting models for the combined photometric and RV data are plotted in Figure~\ref{fig1} and Figure~\ref{fig2}, separately.

As expected, our RV parameters are consistent with those of \citet{knutson2014}, which resulted from the same RV datasets. These RV parameters also agree with the results from \citet{Hartman2011}, though their RV datasets contains four fewer points. As with \citet{knutson2014}, we did not find long-period trend in the RV residuals, so we give the minimum mass of a potential planetary perturber following the convention defined by \citet{wr2007}.

The resulting transit parameters also agree with those from \citet{Hartman2011} except some with slight differences, including a lower impact parameter ($b$), a smaller value for the planet to star radius ratio ($R_{P}/R_{*}$), a  shorter ingress/egress duration ($\tau$), a shorter total duration ($T_{14}$) and a larger Inclination ($i$). Comparing to the published work, which was based on only one full-transit light curve, our results are more robust, as a consequence of being based on the seven complete light curves.

Our stellar parameters show agreement with those of \citet{Hartman2011}, which were chosen as the spectroscopic priors for the global fit in advance. And finally, the planetary parameters of HAT-P-33b calculated based on the derived RV, transit and stellar parameters also agree well with those in \citet{Hartman2011}.

\subsection{Mid-Transit Times} \label{sec4.2}

\begin{figure*}
\centering
\epsscale{0.85}
\plotone{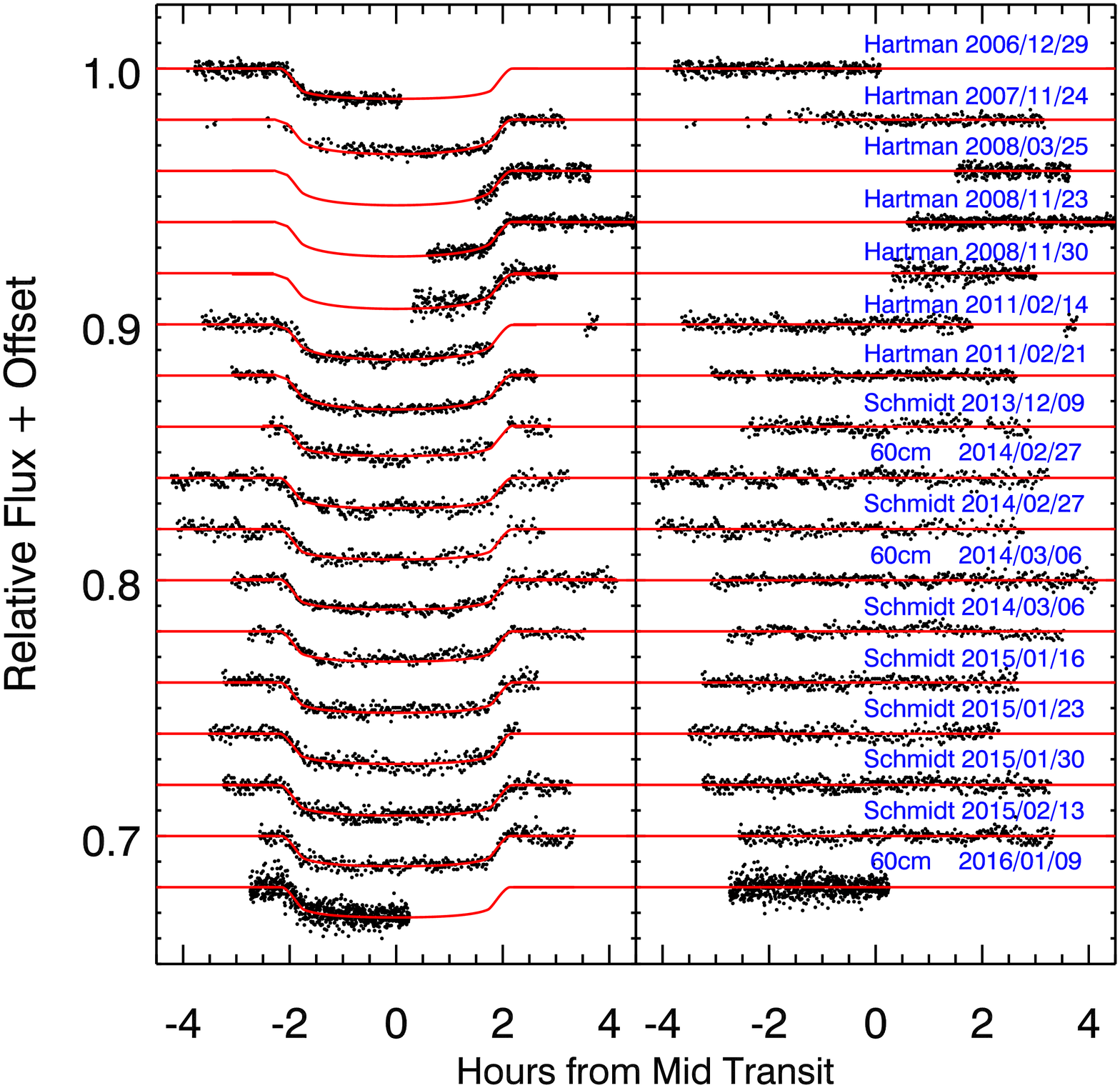}
\caption{Seventeen transit light curves for HAT-P-33 obtained by \citet{Hartman2011} and this work, with which we can estimate the mid-transit times through the separate fits (see \S \ref{sec3}). The resulting best-fitting model for each light curve is shown by the solid red line, with residuals on the right. Both light curves and residuals are displaced vertically for clarity. For more details of our light curves, see Table~\ref{table1}.
\label{fig3}}
\end{figure*}

In order to revisit the orbital ephemeris and seek TTV signals for the HAT-P-33b system, we acquired accurate mid-transit times ($T_{\rm c}$) through separately fitting each light curve. The best-fitting models are shown in Figure~\ref{fig3} and the resulting mid-transit times are listed in Table~\ref{table4}, with uncertainties obtained with the DE-MC method. We fitted obtained transit times with a linear function of transit epoch number ($N$),

\setlength{\tabcolsep}{1.1pt}
\begin{deluxetable}{ccccc}
\tablewidth{0pt}
\tablecaption{Mid-transit times for HAT-P-33b}
\tablehead{
\colhead{Epoch}\tablenotemark{a} &  \colhead{Telescope}\tablenotemark{b} &\colhead{$T_{\rm c}$} & \colhead{$\sigma_{T_{\rm c}}$} & \colhead{$O-C$}  \\
\colhead{}   &\colhead{} &  \colhead{(${\rm BJD_{TDB}}$)} & \colhead{(second)} &  \colhead{(second)}  \\
}
\startdata
 -557  &FLWO &  2454099.85310{ } &  { } 41.47  &{ }   -192.40    \\
 -462  &  FLWO& 2454429.93117{ } &  { } 47.52  &{ }{ }   64.29    \\
 -427  &FLWO &  2454551.53949{ } &  { } 50.11  &  { }{ }  211.56    \\
 -357  & FLWO& 2454794.75180{ } &  { } 31.70  &{ }{ }    112.99    \\
 -355  &FLWO & 2454801.69988{ } &  { } 63.94  & { }{ }   56.88    \\
 -123   & FLWO & 2455607.77606{ } &  { } 51.84  & { }  -112.18   \\
 -121   & FLWO &  2455614.72513{ } &  { } 31.24  &  { } -101.77    \\
  173  & Schmidt & 2456636.22181{ } &  { } 62.21  &  { }{ }  -5.85    \\
  196  & $60\,{\rm cm}$ & 2456716.13465{ } &  { } 59.62  &  { }{ }  -12.65    \\
  196  & Schmidt & 2456716.13498{ } &  { } 56.16  &  { }{ }  15.86    \\
  198  & $60\,{\rm cm}$ & 2456723.08352{ } &  { } 40.15  &  { } -19.52    \\
  198  & Schmidt & 2456723.08417{ } &  { } 50.52  &  { }{ }  36.64    \\
  289  & Schmidt & 2457039.26162{ } &  { } 39.74  &  { }{ }  58.23    \\
  291  & Schmidt & 2457046.20981{ } &  { } 39.74  &  { }{ }  -7.39   \\
  293  &Schmidt &  2457053.15793{ } &  { } 40.61  &  { } -79.05    \\
  297  &Schmidt &  2457067.05749{ } &  { } 37.15  &  { }{ }  64.46   \\
  392  & $60\,{\rm cm}$ & 2457397.13183{ } &  { } 57.02  &  { } -1.12    \\
\enddata
\tablenotetext{a}{The first seven time points are obtained from the published light curves \citep{Hartman2011} through separate fits, the others are from our photometric data. As mentioned above, the epochs (239, 241) were followed by two telescopes simultaneously.}
\tablenotetext{b}{For more information about the FLWO telescope, see \citet{Hartman2011}.}
\label{table4}
\end{deluxetable}

\begin{equation} \label{eq1}
T_{\rm c}[N]=T_{\rm c}[0]+NP,
\end{equation}
where $P$ is the planetary orbital period, $T_{\rm c}[0]$ represents the zero epoch. The best-fitting values are
\begin{equation} \label{eq2}
T_{\rm c}[0]= 2456035.137750 \pm 0.000272\,[\rm{BJD_{TDB}}],
\end{equation}
and
\begin{equation} \label{eq3}
P=3.47447472 \pm  0.00000088 [{\rm days}].
\end{equation}
Our orbital ephemeris agree well with the result from \citet{Hartman2011}.

\begin{figure*}
\centering
\epsscale{1.0}
\plotone{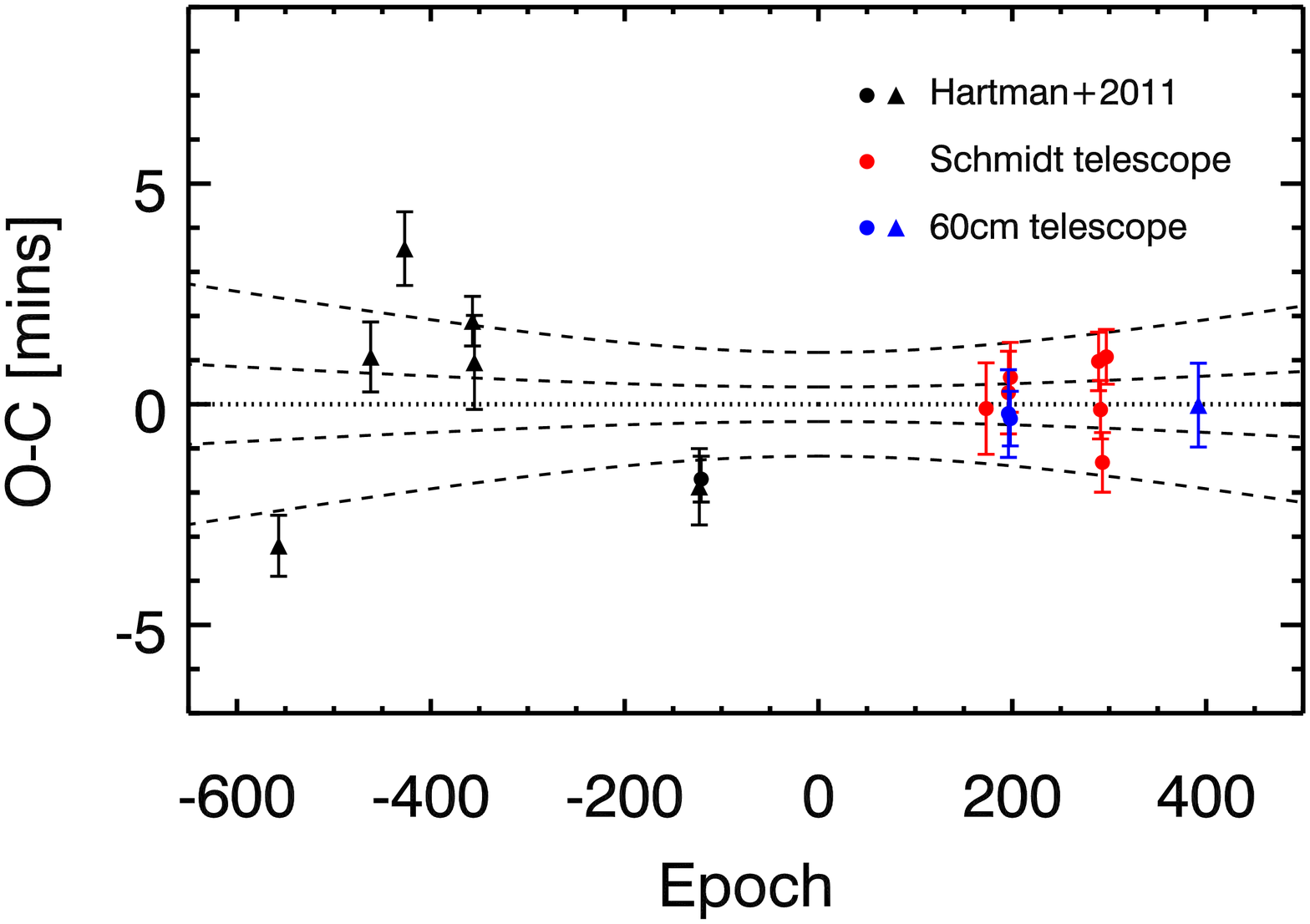}
\caption{The points indicate the residuals of mid-transit times for HAT-P-33b from our linear orbital ephemeris (see eq.\ref{eq1}-\ref{eq3}), which is shown by the dotted line in the figure. The dashed lines indicate the propagation of $\pm1\,\sigma$ and $\pm3\,\sigma$ errors of the orbital ephemeris. We use different colors to distinguish the transits obtained by diverse groups or telescopes. The filled circles mark the full transits and the triangles represent the partial transits. As you can see, the black points are calculated from the published data \citep{Hartman2011} which has only one full transit. The red and blue points are obtained from our photometric data, which have nine full transits and one partial transit.   
\label{fig4}}
\end{figure*}

To get conservative uncertainty estimates for a more reliable future observation schedule, the uncertainties for the mid-transit times during the fitting were rescaled through a common factor to get $\chi^2/N_{\rm dof}= 1$. However, the uncertainties of mid-transit times listed in Table~\ref{table4} were not rescaled in this way, nor were the error bars plotted in Figure~\ref{fig4}.

Figure~\ref{fig4} displays the deviations of mid-transit times from the linear orbital ephemeris (eq.\ref{eq1}-\ref{eq3}), with an RMS of $93.68\,{\rm s}$. This value is largely affected by the mid-transit times derived from the published light curves, which gives an RMS of $144.12\,{\rm s}$ over a $4\,{\rm year}$ time span. As a contrast, the RMS of mid-transit times derived from our data is only $41.87\,{\rm s}$ within a time span of $3\,{\rm year}$. The large deviation of mid-transit times from the published light curves may be mainly caused by their incomplete coverages. In total, most of the mid-transit times are in the $\pm3\,\sigma$ errors range of the orbital ephemeris. Especially for the mid-transit times resulted from our photometric data, which are very consistent with the $\pm1\,\sigma$ errors.

\subsection{Limits On Additional Perturbers} \label{sec4.3}

\begin{figure*}
\centering
\epsscale{1.0}
\plotone{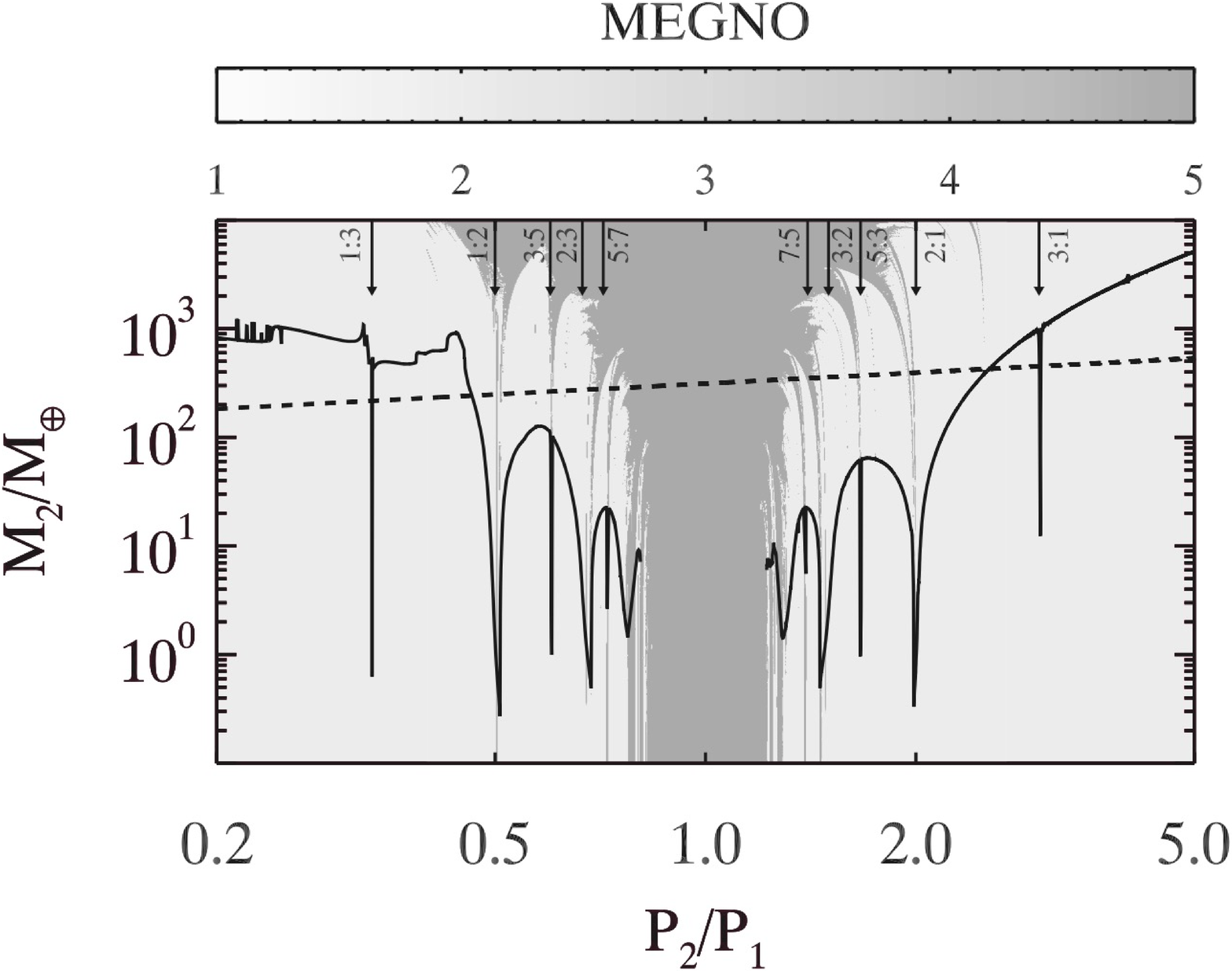}
\caption{The upper mass limits on a potential additional perturber vs. the period ratio of the perturber ($P_2$) and HAT-P33b ($P_1$). The black dashed line shows the loose mass limits of a potential perturber based on the RV residuals (RMS=$47.5\,\rm{m\,s^{-1}}$), from which we can only rule out the existence of a perturber with mass comparable to Jupiter. However, the constraints from TTVs (RMS=$93.68\,\rm{s}$) are much tighter, especially near the the low-order mean-motion resonances (see the vertical arrows) that even a perturber with mass similar to Earth can been excluded. The color coding indicates overall system stability. For values larger than 5, the system is strongly chaotic and hence likely to be unstable. In general, chaotic regions are mainly because of two-body mean-motion resonances (indicated by vertical arrows). When the perturber is in the vicinity to the transiting planet then strong mutual interactions render the system to be unstable resulting in close-encounters or ejections of one or both planets.
\label{fig5}}
\end{figure*}

Although neither significant TTVs nor a residual RV signal were found, we can place the upper mass limits of a potential close-in perturbing planet in the HAT-P-33 system. The results are shown in Figure~\ref{fig5}.

The host star (HAT-P-33) is an active late-F dwarf \citep{Hartman2011}, which has a large RV uncertainty (RMS=$47.5\,\rm{m\,s^{-1}}$). The mass limits based on the RV residuals following the convention in \citet{wr2007} is thus very loose, as indicated by the black dashed line in Figure~\ref{fig5}, which can only exclude a perturber with a mass larger than $0.6\,{\rm \mj}$ near the 1:5 resonances (0.69-day orbit) or $2.0\,{\rm \mj}$ near the 5:1 resonances (17.37-day orbit).

Fortunately, TTV measurements are less sensitive to the stellar activity than are Doppler measurements.  We made use of the MERCURY6 orbit integration package \citet{cham1999} to place upper mass limits of a potential perturbing body. The TTV data exhibited a RMS scatter of $93.68\,{\rm s}$. 

In our simulations, we assumed that the orbits for both the known hot Jupiter and a potential perturber are coplanar and circular, which would gives the most conservative estimate of upper mass limits of the potential perturber \citep{bean2009, fukui2011}. The arguments of periastron, $\omega$, the ascending nodes, $\Omega$, and the initial mean anomalies, $M_{0}$, of the known hot Jupiter and a potential perturber are fixed to $\omega=88^{\circ}$ (from Table~\ref{table3}), $\Omega=270^{\circ}$, and $M_{0}=0^{\circ}$. We explored the mass space of the potential perturber for both interior and exterior orbits with the orbital period ratio from 1/5 to 5 times (0.69 to $17.37\,{\rm days}$) that of HAT-P-33b, which is equivalent to a semi-major axis range from 0.017 to $0.154\,{\rm AU}$. 
We incremented the perturber's semi-major axis by $0.00001\,{\rm AU}$. The resolution is enough to depict the constraints on the perturber mass in the resonant configurations, that the TTV signals are significantly sensitive to \citep{ag2007, holman2005}. In each increment of $a$, we obtained the upper mass limit of the potential perturber by iterative linear interpolation with an initial mass of $1.0\, {\rm M_\oplus}$ and a convergence tolerance of $1.0\,{\rm s}$ for the TTVs.

Comparing to the loose limits by RV data, the mass limits from our TTV measurements are much tighter near the low-order mean-motion resonances, as illustrated by the black solid line in Figure~\ref{fig5}. We can exclude the existence of a perturber with mass larger than 0.6, 0.3, 0.5, 0.5, and $0.3\,{\rm M_\oplus}$ near the 1:3, 1:2, 2:3, 3:2, and 2:1 resonances, respectively.

In Figure~\ref{fig5}, we also present the the dynamical stability in the hypothetical three-body system through Mean Exponential Growth of Nearby Orbits (MEGNO) Index \citep{go2001, ci2003, hi2010}. The resulting dynamical stability map agrees well with that obtained by the analytic method described in \citet{barnes2006}.

\section{SUMMARY AND CONCLUSIONS} \label{sec5}

We initiated the transiting exoplanet monitoring project (TEMP) to study the known exoplanets in great detail, with specific goals of obtaining a better grasp of planetary interior structures, formation and evolution. 

One of the initial targets for TEMP, HAT-P-33b, has been observed by two telescopes from 2013 December to 2016 January. In total, we obtained ten light curves of eight different transit events, thereby substantially enriching the photometric database of HAT-P-33b.

To revisit the system parameters of HAT-P-33b, we have performed a global fit based on our new light curves and the expanded RV data \citep{ knutson2014}. Though most of the results agree well with those from the published work \citep{Hartman2011, knutson2014}, some slight discrepancies still exist in the transit parameters. 

We also separately conducted fits for the seventeen light curves to obtain precise mid-transit times. Along with these, we revisited the orbital ephemeris for HAT-P-33b, which agrees well with that in \citet{Hartman2011}.

Though no substantial TTV signal has been found from the linear orbital ephemeris of HAT-P-33b, we can constrain the upper mass limits of a potential close-in perturbing planet based on the measured TTVs with an  RMS scatter of $93.68\,\rm{s}$. The restriction is much stronger near the low-order mean-motion resonances. We can exclude the existence of a planet with mass larger than 0.6, 0.3, 0.5, 0.5, and $0.3\,{\rm M_\oplus}$ near the 1:3, 1:2, 2:3, 3:2, and 2:1 resonances, respectively. However, we still cannot rule out the existence of additional close-in planets in the non-resonant area. Whether additional planets frequently exist in the nearby non-resonant area of hot Jupiters is an open question. Further work is needed to answer this question, and hence to better reveal the nature of planetary formation and evolution.

\acknowledgments

This research is supported by the Strategic Priority Research Program: The Emergence of Cosmological Structures of the Chinese Academy of Sciences (Grant No. XDB09000000); the National Basic Research Program of China (Nos. 2013CB834900, 2014CB845704, 2013CB834902, and 2014CB845702); the National Natural Science Foundation of China (under grant Nos. 11333002, 11433005, 11373033, 11503009, 11003010, 11373035, 11203034, 11203031, 11303038, 11303043, 11073032, 11003021, and 11173016); the Main Direction Program of Knowledge Innovation of Chinese Academy of Sciences (No. KJCX2-EW-T06); Japan Society for Promotion of Science (JSPS) KAKENHI Grant Numbers JP25247026. Songhu Wang gratefully acknowledges the award of a Heising-Simons 51 Pegasi Postdoctoral Fellowship. Tobias C. Hinse acknowledges KASI research grant 2016-1-832-01. Numerical computations were partly carried out using the SFI/HEA Irish Center for High-End Computing (ICHEC) and the 3rd generation Polaris High-Performance Computing cluster at KASI/South Korea. Research at the Armagh Observatory is funded by the Department of Culture, Arts \& Leisure (DCAL). 

\vspace{5mm}

\facilities{Beijing:Schmidt, Beijing:0.6m}

\software{SExtractor \citep{Bertin1996}, EXOFAST \citep{eastman2013}}

\clearpage

\end{document}